# Mikhail Vasil'evich Lomonosov

## *Oration on the Origin of Light Presenting A New Theory of Color.* [1]

*delivered in a public meeting of the Imperial Academy of Sciences,*

*on July 1, 1756, by Mikhail Lomonosov*

[ *Oratorio De Origine Lucis Sistens Novam Theoriam Colorum* ]

[ *Слово о происхождении света* ]

The trial of nature is challenging, listeners, yet it is pleasant, beneficial, and sacred. The deeper one's understanding of its mysteries, the greater joy the heart feels. As our reasoning delves further into its essence, it gathers abundant fruits for life's needs. The more profoundly we penetrate the marvels of creation, the clearer the incomprehensible architect of existence reveals Himself. The visible world, with its omnipotence, grandeur, and wisdom, is the primary, universal, truthful, and unceasing preacher. The heavens declare the glory of God. He has set His tabernacle in the sun, revealing the radiance of His divinity more clearly than in other creatures. Through the immeasurable expanse of the universe, it shines incessantly beyond the farthest planets, spreading a multitude of rays surpassing human imagination. These perpetual and lightning-fast messengers, gentle and favorable in nature, illuminate, warm, and animate creation, arousing a divine imagination not only in the human mind but also seemingly in dumb animals. How should those who gaze with curious eyes into the inner sanctuary of nature contemplate the boundless ocean of light? Numerous works, shared across different nations and ages, attest to their tireless efforts and contributions, overcoming many obstacles and easing the labors of future generations. They dispelled dark clouds and penetrated deep into the clear sky. But just as the physical eye cannot directly gaze at the sun, so too does the vision of reasoning dull when probing the origins of light and its division into various colors. Shall we abandon hope? Retreat from the task? Succumb to despair over our achievements? Never! Shall we appear indifferent and unworthy of the efforts of those who tested the boundaries of nature? Let us assess the great assembly of materials they have gathered for this task, or, as ancient tales tell of giants, the vast mountain they have raised, daring to approach the source of such radiance and the splendor of colors. Let us ascend fearlessly beyond them, stand on their sturdy shoulders, and, rising above all preconceived notions, direct our minds and reasoning, as far as possible, to explore the causes of light and the separation of its colors.

At the outset of this endeavor, let us examine the foundation of this great structure, erected by numerous, sometimes agreeing, sometimes conflicting builders; and where it is disorderly and unstable; and we will endeavor to rectify and strengthen it to the best of our ability with the tools of our own thoughts. After that, we will begin to construct our own system.

---

[1] Translated from Latin and Russian versions (Refs. [1]) by V.Shiltsev, footnotes and commentary by V.Shiltsev.

Colors originate from light; therefore, we must first examine its cause, nature, and properties in general, and then investigate their origin. Bypassing the hidden views of the ancients, I turn to the opinions of our time, enlightened by the clearest physical knowledge. Among them, two are paramount: the first, Cartesian, confirmed and elucidated by Huygens[2], and the second, which was initiated by Gassendi and became of importance through Newton's agreement and interpretation[3]. The difference between the two lies in their respective motions. In both, the finest, liquid, imperceptible matter is postulated. But in Newton's theory, the movement is assumed to flow and spread in all directions from luminous bodies, resembling a river; in Descartes', it is assumed to be perpetually undulating without flowing. Which of these opinions is correct and sufficient for explaining the properties of light and colors, let us ponder with attention and caution.

For a clear and detailed understanding, it is necessary to consider all possible forms of motion in general. Thus, assuming the existence of a liquid, finest, and intangible matter of light, which no one doubts today, we find three possible motions within it, which may or may not exist — this will be determined later. The first motion may be flowing or passing, as Gassendi and Newton suggest, where the ether (the matter of light, referred to by ancient and many modern scholars) flows constantly from the sun and other great and small luminous bodies in all directions, resembling a river. The second motion may be undulating within the ether, according to Descartes and Huygens, where it acts like very fine and frequent waves in all directions from the sun, extending across the vast expanse of universal space filled with matter, much like still water spreading waves in parallel circles in all directions from a fallen stone, without any flowing movement of its own. The third motion may be circular, where each insensible particle composing the ether revolves around its center or axis. These three possible motions of the ether may or may not truly exist within it and produce light and colors — this will be thoroughly and attentively investigated.

The opinion that attributes the cause of light to the flowing motion of the ether is merely an arbitrary assumption, lacking any foundation or evidence. Only two circumstances lend some semblance of probability: first, the laws of refraction devised by Newton, and second, the perceptible time it takes for light to reach us from the sun. However, these rules are based on a similarly arbitrary assumption about the attractive force of bodies, which eminent physicists rightfully reject as a hidden quality revived from the ancient Aristotelian school, bordering on absurdity. Therefore, although they demonstrate the author's ingenuity sufficiently, his opinions are by no means affirmed. The perceptible but very brief time it takes for light to travel from the sun to the earth even less confirms the flowing motion of the ether than the lingering of sound in the air after a striking blow at a considerable distance confirms the flow of air. If someone claims that light from the sun is carried by the flow of the ether like a river because there is a perceptible time interval when light reaches our vision, they must conclude by the same reasoning that air flows in all directions from ringing strings at the same speed as sound travels to the ear. However, I imagine the speed of a strong wind, when air rushes at 60 feet per second, causing great waves on the waters and uprooting trees with their roots, and I reason that if air moved as swiftly by passing flow from strings as sound does, that is, more than a thousand feet per second, then mountains would be displaced from their places by such music.

---

[2] The theory of light by R. Descartes (Cartesian) was presented in his famous *Discours de la méthode* (Discourse on the Method. Leiden, 1737), - in the appendices to this 'Discourse' - '*Dioptrique*' and '*Météores*'. The wave theory of light by C. Huygens (Huygens), substantially different from the views of Descartes, was published in the book *Traité de la lumière* (Treatise on Light. Leiden, 1690).

[3] P. Gassendi outlined his views on the nature of light in *Syntagma philosophicum* (Philosophical Compendium), printed in the first volume of his works (Lyon, 1658). Gassendi regarded light as a direct emanation of matter from bodies. The theory of light and colors by I. Newton is presented in his book *Optics, or a treatise of the reflexions, refractions, inflexions and colours of light* (London, 1704; Latin edition, London, 1719).

Although both of these mentioned conjectures, employed to support the establishment of that opinion, can serve as little probable evidence, let us concede for the time being and, assuming that light from the sun spreads in all directions by the flow of the ether, let us see what follows.

From mechanical laws, it is sufficiently proven, confirmed by everyday experiences, and universally accepted that the smaller and lighter an object is, the less resistance it offers to the driving force, and the less tendency it receives; likewise, the greater the opposing resistance it encounters, the sooner the motion of that object ceases. For example, if someone were to hurl a grain of sand from a sling, would it fly with the same speed and over the same distance as a stone propelled by the force of a human hand? Now, consider how much smaller and lighter is a single particle of the ether that constitutes light? And how vast is the distance from us to the sun? And what sort of motion can be imagined for the ether according to the aforementioned opinion? And what resistance could be stronger than the gravitational pull toward the sun, which not only affects our Earth but also other large bodies, deflecting them from their straight paths? Can we attribute the origin of light to the flowing motion of the ether in such circumstances?

Let us expose a small, black, and opaque grain of sand to sunlight for twelve hours. During this time, rays from the entire visible solar hemisphere will continuously converge upon it, forming a vast cone, with the sun at its base and the grain at its sharp end. The cubic volume of this conical space, according to calculations, contains around seven hundred and twenty million cubic earth diameters. Every eight minutes, the spread of light from the sun to the earth occurs; therefore, in twelve hours, approximately eight hundred and forty million cubic earth diameters of etheric matter will flow from the sun to that grain. If we take the grain exposed to sunlight and place it in a small, dark, and cold chamber, the warmth acquired from the sun will vanish instantly, and not the slightest trace of light will be found. Even if someone were to repeat this experiment for a whole year or their entire lifetime, the blackness of the grain would persist, emitting no light in the darkness. The black substances do not repel the rays that fall upon them, nor do they allow them to pass through. Tell me, advocates of the theory of light produced by the flowing motion of matter, where does it disappear in this case? You cannot say otherwise than it gathers in the grain and remains there entirely. But is it possible for such a quantity of matter to fit within it? I know that you divide the matter of light into exceedingly small particles and scatter it sparsely throughout universal space, claiming that the entire quantity can condense and fit into the sparse cavities of a single grain of sand. Although your division lacks any foundation or evidence, I concede to you under the condition that I am also allowed, by your own reasoning, to divide matter into equally small parts. You cannot deny me that. Therefore, I divide the surface of the black and opaque grain into numerous millions of parts, each illuminated by the entire visible solar hemisphere; to each of these parts, the same tremendous amount of etheric matter flows, fits, and remains. Where will you show me so much space? Will you divide matter even finer? But in the same manner, I also have the right to divide my particles on the surface of the grain and demand the same amount of light for each. See how burdensome your arbitrary opinion is!

Yet you may still say that while we see inconveniences, we do not see impossibility, which alone could prove the untenability of our argument. My reply: inconvenience often dwells alongside impossibility, which I have happened to discover within your argument with more than one argument.

What is harder than a diamond among known substances? What is more transparent? Hardness requires a considerable amount of matter and close cavities — transparency barely allows for any matter at all, assuming that rays extend through the flowing motion of etheric matter. For from every point on its surface and from every point within its entire body, rays travel in straight lines to every other point on its surface and within its entire body. Consequently, straight-line cavities extend inward throughout the entire diamond. Assuming this, the diamond must not only consist of rare and loose matter, but its entirety must also be hollow. From hardness, its composition must consist of closely connected particles; from transparency, it follows not only looseness but also nearly entirely hollow, surrounded by a thin shell. These consequences mutually contradict each other; therefore, the arbitrary assertion that light from the sun extends through the flowing motion of etheric matter is unjust.

Let us further assume that light extends from the sun and other luminous bodies through the flowing motion of ether. New impossibilities, new paradoxical conclusions will follow. In a transparent diamond, straight-line cavities extend throughout its entire body, through which light matter passes, as previously demonstrated. Light is transmitted from one side to another without obstruction. Let us place a diamond between two candles. The rays from both sides will pass through the diamond with equal force, and one candle will be as distinctly visible from one side as the other candle from the opposite side. What do we make of this? Shall we discard mechanics? Shall we suppose that when equal forces and equal quantities of liquid matter meet in a narrow cavity from both sides, as they should through the diamond, they do not encounter each other and do not impede one another?

But is that all? Through all the cavities of a diamond placed among thousands of burning candles, how many opposing and intersecting flows of light matter must there be, through countless angles of inclination? Yet there is no obstruction, no disturbance in the rays, even in the slightest! Where are the just logical conclusions? Where are the unbroken laws of motion?

These refutations might suffice; however, to further undermine the credibility of this theory, I propose to consider the following: is it possible in nature for the same thing to be larger than itself? Immutable mathematical laws assert that the same thing is always equal to itself in magnitude. Anything to the contrary is unjust and contrary to common sense and human reason. However, from the arbitrary assumptions of Gassendi and Newton, this conclusion undoubtedly follows. Sun rays are reflected from the sides of glass prisms with such intensity that objects placed there appear as distinct as if directly observed. This suggests that all the rays are repelled from one surface and only a few pass through it. Conversely, objects can be clearly seen through the same surface as if directly exposed to sight. From this, it indisputably follows that all the sun rays pass through that surface, and only a few are repelled. Does this not imply that according to this theory, as many rays are repelled from one surface as fall upon it, and as many pass through it, meaning that the matter of sunlight is doubled? Now we must hold one of two assertions: either the theory of light rays extending through the flowing motion of ether is false, or it is true, and we must believe that the same thing can be larger than itself at the same time.

Having considered the impossibility of this motion of etheric matter, let us turn to the second, namely, the rotational motion, and see if it can be the cause of light.

In my "Discourse on the Cause of Heat and Cold,"[4] I have demonstrated that heat originates from the rotational motion of particles constituting the bodies themselves. Although the unfairness of any objections has been clearly shown, I shall briefly reassert it with new arguments from practice.

When iron is forged, it heats up, its own matter contracts denser, and extraneous matter escapes, clearly proving that the external matter, diminishing, does not cool it—its own matter, agitated by friction and the motion of particles, gets hotter.

When copper or another metal is dissolved in strong spirit[5] or soaked in lime water, heat is spontaneously generated in them without any external heating. According to proponents of heat-producing matter, it must gather from nearby bodies, and therefore, these bodies must cool down. But this is contrary to all experiments. Therefore, the accepted notion of heat-producing matter contains both equilibrium and lack thereof: it contains equilibrium when heat exits a warm body into a cold one, heating it and itself cooling down to an equal degree of heat; it does not contain equilibrium when lime is heated without cooling the surrounding objects—a clear paradox.

---

[4] Lomonosov's proposition regarding the rotational motion of particles as the cause of heat is contained in § 11 of his "Reflections on the Cause of Heat and Cold" (see vol. 2 of PSS [2], pp. 20-21).
[5] Concentrated nitric acid

Lead in boiling water, no matter how long it remains, does not absorb any more heat, as indicated by a thermometer. According to the supporters of heat-producing matter, it emerges from fire into heated matter, enters insensitive cavities, and fills them up according to their size. However, the same lead outside water absorbs significantly more heat, melts, ignites, and turns into glass. Here, according to the matter exiting and entering, it must be concluded that the same lead outside water has larger cavities than inside them, and it becomes uneven and unlike itself at the same time it remains lead.

Red-hot iron is extinguished by boiling water. Therefore, according to those who consider the cause of heat and cold to be in fiery matter transferred from one body to another, it emerges from iron into boiling water. But from well-known experiments and indisputable conclusions, it is evident that water, when boiling, cannot be hotter. Therefore, according to the same theory, it does not absorb any more heat from within. See—an obvious paradox! At the same time, from the same iron, water absorbs heat-producing matter and does not absorb it.

From animals, warmth constantly extends and heats nearby objects. Many of them never partake of warm food. Advocates and defenders of the heat-generating substance, explain: by what means does it enter animals insensitively, sensitively exit? Is it, when it enters, sometimes cold? That is, warmth can be cold, just as light can be dark, dryness can be wet, hardness can be soft, roundness can be angular!

All these difficulties, or, better to say, impossibilities, will be eliminated when we assume that warmth consists of the circular motion of insensible particles, constituting bodies. There will be no need for the strange and incomprehensible passage of a certain heat-generating substance from body to body, which is not only unsupported by evidence but can also be interpreted less clearly. Circular motion of particles is sufficient to explain and demonstrate all the properties of warmth. For further assurance on this matter, I refer hunters to my Discourse on the causes of heat and cold, as well as to the responses to critical arguments against it.

Now, let us consider whether the circular motion of ether particles can be the cause of light.

Though the sun shines and warms, there are many cases where there is no light even with great heat, and where clear light is not accompanied by warmth. Iron taken from a furnace, when already extinguished, does not emit any light in darkness, but it contains such heat within itself that it causes water to boil, ignites wood, and melts lead and tin. On the contrary, rays collected by a burning mirror, though repelled by the full moon, shine very vividly and clearly, but they do not produce sensitive warmth. I do not mention the electric light of phosphorus and others, which emit light without heat in darkness. So, when fire can exist without light and light without fire, it follows that both arise from different causes. Ether communicates light and warmth from the sun to earthly bodies. Therefore, it must be concluded that both are produced by the same substance, but by different motions. The impossibility of current motion is proven; circular motion is the cause of light and warmth. Therefore, when ether in earthly bodies produces warmth, that is, the circular motion of particles, it must itself possess it. Therefore, when ether cannot have current motion, and circular motion is the cause of warmth without light, then only one third remains, the wavering motion of ether, which must be the cause of light.

Although this is already sufficiently proven, let us still consider: first, whether there are any absurd consequences of the wavering motion of light similar to those produced from the notion of the current motion of ether; second, whether different properties of light can be interpreted.

As for the first, we have a clear example in the wavering motion of air, through which sound extends from place to place. Everyone can easily imagine how many different sounds there are, just by thinking about different musical tones, varying loudness, sounds from different instruments, as well as the voices of birds and other animals; also, about thunder, ringing, banging, cracking, whistling, squeaking, creaking, murmuring, and their various intensities and elevations, as well as different articulations of letters in different languages. All these countless differences in sound extend in a straight line, intersecting each other not only at every possible angle but also directly meeting, without destroying one another. Standing near ringing strings, I hear in one direction the song of a nightingale, in another the voices and speech of singers; there is the sound of ringing bells, elsewhere the sound of horse hooves; all these sounds come to my ear and to many others, and the one we pay more attention to, the clearer we hear it. Thus, we have evidence that nature employs the wavering motion of liquid bodies, such as air, for great and varied purposes. Similarly, having presented the demonstrated impossibility of current etheric motion, without doubt, we must accept its wavering motion as the cause of light, for there is no absurd consequence from the aforementioned wavering motion. There is no need to cram into one particle a substance that occupies space between it and the sun, which is of such vast extent. There is no need for a diamond to be anything more than a thin, fragile shell. There is no need to accept other absurd opinions.

Secondly, the convenience of this system, which undoubtedly serves the clear interpretation of the actions and circumstances of light, confirms various motions as the causes of warmth and light.

As shown above, the rays from the lunar semicircle, constricted by a burning mirror, do not show sensitive warmth, they have light, barely tolerable to the eyes. This wonderful property will be clearly and understandably explained by the aforementioned positions. Etheric matter between the sun and the moon moves with its particles in a wavering and circular motion. The circular motion warms the surface of the moon and dulls it; the wavering motion, which serves not for warming but for illuminating, loses less of its power, so that the repelled rays from our earth reach the moon and return from it again, showing part of its dark side soon after the full moon.

Mercury in a glass vessel, containing no air, produces light without warmth when falling in small drops. It is well known to all knowledgeable people that a round liquid drop, after striking a solid body, trembles, compresses, and expands, thus inducing the ether into a trembling motion, which produces light. Thus, phosphorus and other similar substances emit light without heat. Explanation of these phenomena can now be sufficiently brief.

Next, it is in order to announce my opinion on the cause of colors and to prove it by probability. But before I present it, I will show the basis, which is still unknown in all of physics to this day, and not only does it lack interpretation but also a name, yet it is so important and universal in all of nature that in the production of properties originating from insensible particles, it occupies the foremost place. I call it the *combination of particles*. The strength of this basis depends on the similarity and dissimilarity of the surfaces of particles of the same and different kinds of original matter, constituting bodies.

Imagine the space of the universe as composed of tiny spheres, insensitive yet varying in size. These spheres form a surface filled with frequent and small irregularities, akin to the cogs on wheels, which can interlock with one another. From mechanics, it is known that wheels interlock and move together when their cogs are of equal size and in the same arrangement, meshing harmoniously. Conversely, if the sizes and arrangements differ, they do not interlock or move together. I find this analogy in the insensible primary particles constituting all bodies. These particles, designed and regulated by the wise architect and omnipotent mechanic, adhere to immutable natural laws. I term particles that interlock and move together as *compatible*, and those that do not as *incompatible*.

By envisioning this foundation, one can easily comprehend all the sensory perceptions and other marvelous phenomena and changes occurring in nature. The vital fluids within the nerves announce changes at their ends through such movement, engaging with particles of external bodies in contact. This occurs seamlessly over time, ensuring continuous alignment of particles along the nerve from end to brain. According to mechanical laws, many thousands of such spheres or wheels, when continuously interlocked, must rotate with a force applied externally, stop when halted, and collectively increase or decrease the speed of motion.

In this manner, acidic matter contained within the nerves of the tongue engages with acidic particles placed on the tongue, producing motion changes and conveying them to the brain, thus giving rise to the sense of taste. Similarly, chemical solutions, descents, boilings, and the ascent of liquid substances in narrow tubes occur. Through this instrument, electric force operates, clearly presented, explained, and demonstrated without the need for the inexplicable influx and efflux of miraculous substances without any cause. Consider that through the friction of glass, a circular motion of its particles is produced within the ether, opposite in speed or direction to the movement of other ether. This motion extends from the surface of the glass into convenient water-filled or metallic tubes. Here, there is no need for the incomprehensible flow of ether particles; only a slight rotation suffices. Unlike the case where ether flows through a very long and convoluted wire without any repulsion or collision, retaining its movement in myriad directions for a long time, the discomforts of such movement are eliminated by the rotational motion of interconnected ether particles. As such, electric sparks, sensations of illness, thunderous strikes, and other phenomena and properties, according to previous interpretations, appear even more wondrous than clear. Through this system of particle combination, the mechanical explanation becomes easily understandable. However, brevity does not permit further elucidation, and the splendid colors from thunderous electric clouds beckon my words to return to them.

All the aforementioned etheric particles, of which there are countless, I classify into three categories of varying sizes, all of which are spherical. The particles of the first category are the largest, in continuous mutual contact and in a square arrangement. Therefore, assuming a cubic body against a sphere of the same diameter, there remains almost as much space between these particles as the spheres occupy. In these gaps, I posit etheric particles of the second category, much smaller than the first, fitting a significant number in each and occupying half the space of these gaps through continuous contact and square arrangement, hence possessing half the matter. Likewise, I infer the existence of the third category of the smallest etheric particles in the gaps of the second category particles. These third-category particles are similarly arranged and, according to the aforementioned geometric size, have a matter quantity ratio to the second category as one to two and to the first category as one to four. I see no reason or need for further division into finer particles. Each of these three categories of etheric particles is compatible with its own kind and incompatible with particles of other categories. Thus, when a particle of the first category rotates, attached to others of its kind, a significant number move at a considerable distance with the force of attachment. Particles of the second and third categories are not involved in this movement. The same applies to the other two categories of particles. In short, two categories of particles can remain stationary when one rotates, and when two rotate, one can remain immobile, just as all three can move or be at rest independently of each other.

Sensitive bodies, according to the division and agreement of the most eminent chemists, consist of primary matters, some acting and some accepting, or principal and subsidiary. In the former, they postulate saline, sulphurous, and mercurial matters; in the latter, pure water and earth. They do not consider ordinary salt, sulphur, and mercury to be the most primary and simple matters, but only borrow their names because of the dominance of those primary materials.

Through many years of conjecture and subsequently through demonstrative experiments, I have firmly established, with sufficient probability, that three types of etheric particles combine with three types of active primary particles, constituting sensitive bodies. Specifically: ether of the first magnitude with saline, of the second magnitude with mercurial, of the third magnitude with sulfurous or combustible primary matter; while with pure earth, water, and air, the combination is dull, weak, and imperfect. Finally, I find that from the first type of ether arises the color red, from the second - yellow, from the third - blue. Other colors arise from the mixture of the first ones.

Having seen the structure of this system, let us examine its motion. When the sun's rays extend light and warmth onto sensitive bodies, then, with a wavering motion, etheric spheres touch and press against their surface, while with a circular motion, they rub against it. Thus, combined etheric particles engage with the corresponding particles of the primary matter constituting the bodies. And when these are inconvenient for circular motion for any reason, then the circular motion of that type of ether is dulled; the wavering motion remains in force. In such circumstances, the following phenomena occur. When the particles of a sensitive body are arranged in such a way that each primary matter has a place on its surface, then all types of etheric particles touch them, lose circular motion through combination, and without it, solar rays produce no colors in the eye, lacking the power to induce circular motion in its constituent parts. Thus, the bodies appear black. Suppose the mixture of a sensitive body is such that none of the dominant primary matters is present on the surface of its mixed parts, but they are surrounded by particles of pure earth or water. Then all types of etheric matter must have weak combination with them, and circular motion hardly encounters any obstacles. Consequently, with a trembling motion at the bottom of the eye, it acts, producing all colors in the visual sensation, and such mixed bodies have a white color.

Then let there be on the surface of mixed matter particles of primary acidic matter; others either absent in the mixture or covered by acidic matter. Then, etheric matter of the first type, lacking circular motion for combination with them, will not produce the sensation of red color in the eye, and only yellow and blue ether, freely acting, will affect the optical nerves on the mercurial and combustible matter, producing the sensation of yellow and blue colors simultaneously, from which such bodies should be green. Similarly, on the surface, mercurial matter alone produces a cherry color, the combustible one - a reddish-yellow color in bodies.

When two matters have a place on the surface of mixed particles, then from acidic and mercurial, only the sensation of blue remains there, from acidic and combustible - yellow, from mercurial and combustible - red, because in the first case there is no combustible matter on the surface to elevate blue ether, in the second there is no mercurial to retain yellow, and in the third there is no acidic matter to elevate red ether.

You can already see the entire system of my views on the origin of colors; it is necessary, finally, to offer proofs and assure that my proposed idea is more than a simple invention or arbitrary assertion.

Firstly, as to the triadic number of colors, numerous optical experiments conducted by the renowned physicist and diligent investigator of color phenomena, Mariotte, assure every free-thinking person of informed thoughts. He attempted not to refute, as some thought, but to correct Newton's theory of light separation by refraction into colors, and only to affirm that in nature there are three, not seven, primary colors.[6]

The nature itself requires particles of different sizes and their aforementioned arrangements, and that demands equal division of them everywhere, so that the proportion of three types of ether is maintained everywhere and so that it does not lose this proportion by any inclination or resistance, and that each type of continuous combination is not deprived. I explain this with a simple and very understandable example. Imagine a space filled with cannonballs, so that no more can fit in it. However, there will be empty spaces between them, which can accommodate a great number of musket bullets. Let the spaces between the bullets be filled with fine shot. In such a state, let the cannonballs, bullets, and shot come into motion in any way imaginable. The cannonballs will remain in equal proportion everywhere; similarly, the bullets will always find their place among the cannonballs in proportion; and there will be an equal amount of shot between the bullets. And thus, continuous contact between the three types of spheres will be maintained. This method, and only this one, is possible to preserve the equal proportion of the mixture of three types of ether everywhere. For if the ether differed in shape or weight, then it would be impossible for it to remain in a uniform mixture everywhere. Look at the movement of air, at sea waves, at the annual and daily flow of the earth, at the orbits of planets and comets; there always remains an equal proportion of ether in their mixture, regardless of their inclination and strength. Each type does not gather in one place, excluding others. And it is impossible for this to happen, given the arrangement described above. In other circumstances, it would be necessary for this to happen.

Nature, above all, is most astonishing in its simplicity, and from a small number of causes it pronounces countless shapes of properties, changes, and phenomena. Why, then, should it have special types of ether for reddish-yellow, for green, for cherry, and other mixed colors, when it can form red-yellow from red and yellow, green from yellow and blue, cherry from red and blue, and other types of mixed colors from other different mixtures? Painters use primary colors; others are made through mixing; can we assume that in nature there are more types of ether material for colors than it requires, always seeking the simplest and shortest paths for its actions?

---

[6] The theory of colors is presented by E. Mariotte in his book *De la nature des couleurs*, Paris, 1681 (On the Nature of Colors. Paris, 1681).

Furthermore, the fact that refracted light through prisms with proper precision shows the triadic number of primitive simple colors is evident in bodies consumed by fire. When a candle, wood, or other body that ignites with lively and free flame burns, we see red fire in the coals, yellow in the flame itself, and blue between the coals and the yellow flame, that is, three primary matter particles, constituting that body, brought into circular motion by the heat of the burning body, move ether of three kinds. In the coals, acidic matter moves the combined ether red; in the flame, mercurial matter - yellow; above the coal, combustible matter - blue, for it, more conveniently and before the mercury in the flame, turns blue ether into circular motion. All this gains greater probability from the following.

Pure double spirit[7] contains most of the combustible matter in it, and besides a little acidity, no one has noticed anything mercurial in it. When ignited, it burns with a blue flame, clearly showing that the combustible primary matter, turning in circular motion in it, turns the third kind of ether, combined with it, and produces the sensation of the blue color. Mineral sulfur, besides combustible matter, contains acidic matter, does not have mercury, and therefore, when ignited, gives a cherry color, which should be the case according to this system. For, when the particles of acidic matter turn, red ether is brought into circular motion, which together with blue is capable of producing the imagination of a cherry color. Mercury primary matter should, according to the above, produce a yellow flame. This is evident from the art of artillerymen, who use antimony - a body abundant in mercurial matter - in festive fires to produce a yellow flame.

Phosphorus, when glowing or igniting, shows a greenish color, which corresponds clearly to its mixture, for phosphorus consists of combustible matter and saline acidity, which is mixed with mercurial matter.

Gold, when after melting cools down and approaches the state of a solid body, then shines with a very pleasant green light. What happens then in its mixture? Acidic matter loses circular motion before all others (because it requires more heat), the other two, combustible and mercurial, still have enough heat to turn the particles in rotation, turn the ether of the second and third kind, and produce the sensation of yellow and blue together, that is, of the green color.

The green color of flame, although observed from many burning bodies, is most prominent from copper. Moreover, it is noteworthy that when it melts, the flame becomes entirely green when fresh cold coals are added. This occurs for the same reason that the greenness of cooling gold arises, namely, the heat of the flame diminishes with the addition of cold coal, the acidic matter of the hot copper loses its circular motion force, and the combustible and mercurial matters move at a sufficient speed due to the weak heat. Thus, without the movement of red ether, yellow and blue together create the sensation of green in vision.

---

[7] Ethyl alcohol

These experiments, confirming my opinion by their agreement, demonstrate the action of primary matters when, in flames, they bring ether into circular motion and produce various colors in vision through their combination. Now, it is necessary to show how it is repelled from the surfaces of illuminated bodies and produces different colors through various combinations. For this, let us first consider the blackness and whiteness of tangible bodies, then we will proceed to colors.

When water boils, it no longer absorbs more heat. Consequently, the combination of its particles with other surrounding matter, brought into circular motion, cannot occur at an equal speed. Thus, ether particles, not having precise combination with water, placed on the surface by the displacement of sensitive bodies, come to vision with the circular motion of all three kinds of ether and excite the sensation of all colors, that is, the color white. But when fire touches white combustible matter, for example, paper or wood, it immediately turns black and into coal. Why does this happen? The water, having been in mixture, is driven away by heat, the active primary matters remaining exposed, holding the ether together from circular motion, which, not reaching our eye, does not produce any color sensation in it, and therefore, blackness is presented to us. From this, it follows that white objects are less, and black ones are more heated by the sun. For all three kinds of ether material, by particles of black bodies, are caught by combination and induce them into circular motion; the opposite happens with white ones.

A large light collecting mirror, coated with black lacquer, produces great light at the focus point, but little heat, clearly showing that the circular motion of ether in black matter has tired, leaving it oscillating freely.

Here you may reasonably ask me why I do not provide one cause for both heat and colors, since they are different phenomena? I answer that the movement producing heat and colors is circular, but the matters are different. The cause of heat is the circular motion of particles constituting sensitive bodies. The cause of colors is the circular motion of ether, which together with heat imparts to earthly bodies from the sun. The affinity of considerable heat and colors is evident from this, but we will see more if we delve further into the nature of both properties. For the present case, a new observation may suffice, namely, that the colors of cold bodies appear livelier to the sight than those of warm ones.

Take two pieces of the same-colored material, especially red, from the same piece. Place one on a hot stone or iron, just enough to avoid ignition; the other - on a cold one, especially in severe winter frosts. You will clearly see that on the cold stone, the material appears noticeably redder than on the hot one. This truth can be ascertained by exchanging parts of the material from the hot stone to the cold one and from the cold one to the hot one, as many times as you like. Other colors do not change so sensitively.

Here it can be clearly seen that in cold bodies, their constituent particles rotate with less circular motion, and the ether is more exalted. And those particles which are not on the surface of the mixture leave free that which does not have combination on the surface; thus, what is separated from others appears clearer. On the contrary, in hot bodies, particles move more quickly; ether particles do not hold them as strongly from circular motion; hence, by their movement, the main color is dazzled and does not come to sight as vividly. I concluded this initially based on my theory, and later found it to be true through art.

Now it's time to look into all three dominions of diverse nature, in order to briefly demonstrate how great the similarity is in the composition of animals, plants, and mineral substances with this system.

From chemical experiments, it is known that in animal mixtures very little free acid is found; hence, there is little greenness in them. Therefore, when parts of animals decay, it is not acidity that follows, but putrefaction. Acidic and combustible by acidity, mercurial primary matter is freed from the mixture by putrefaction. Hence it is evident that acidic primary matter is enclosed in animals by others and produces little sour taste and green color.

On the contrary, in growing things, greenness and acidity abound: in all parts where there is greenness, acidity is also perceptible; in flowers, acidity and greenness are lost. Immature fruits are sour and green, ripe ones are adorned with blue, blush, yellow, or crimson, and receive various kinds of sweetness, which either diminishes or completely suppresses acidity.

When a tree decays or leaves fall from it, then they display a yellow color: through putrefaction, mercurial matter separates from the mixture and scatters into the air. Consequently, the second kind of ether, that is, yellow, does not have combination on their surface, does not lose circular motion, and, reaching our eye, produces the same in the combined mercurial matter in the black membrane at the bottom of the eye and in the optic nerve, exciting the sensation of the yellow color.

In the world of minerals, which have greater relation to chemistry, I could present a great number of examples to establish the truth of my opinion, elucidating the properties and phenomena of colors in various mountainous substances and chemical actions. However, all of them cannot be contained in my present discourse. Therefore, I will present a small part of them.

Water and pure earths and stones have no color other than white, that is, all three kinds of ethers repel them, not taking away their circular motion. This corresponds to what was shown earlier, that they have little combination with ether. On the contrary, black bodies are always composed of many different materials mixed together, and being combined with ethers of all kinds, they hinder their circular motion, without which no sensation of any color can be depicted in the eye.

I cannot fail to mention here the opinion, that is contrary to everyday realities, of those who, assuming the extension of light through ether, attribute blackness to the multitude of holes which they assign to black bodies, and assert that light, entering them, disappears. According to their opinion, the more holes any body has, the blacker it should be, the fewer, the whiter. Therefore, white chalk should be denser than black marble, ground pigments darker than unground ones, to all of which we find the opposite in nature.

A different but corresponding to my theory phenomenon is seen in the making of ink. When its constituent matters are still in separation, their particles move freely in water with circular motion, almost not exalting etheric globules, and hence, their color is of notable blackness. But when combined together into one mixture, then the combined particles become large and are little suited for circular motion; then all three kinds of ether are exalted in circular motion, and not coming to the eye with it, they do not produce any sensations of colors and represent the mixture as black. By adding strong spirit, the ink whitens, because acidity separates the combined matters and thereby gives them greater freedom of movement; from alkaline salt[8], blackness returns to the ink, because acidic matter, having taken its own in the mixture, gives freedom to recombine with the matters constituting the ink.

Such combination into large mixed particles of primary particles, constituting bodies, occurs in all chemical descents, when, separating from liquid solutions, dissolved matters combine into coarse particles, settle to the bottom, and produce various colors, depending on which matters occupy their surface in greater quantity.

Hence it happens that the most acidic mineral liquid matters do not have a green color, because they move freely in water and etheric red does not exalt in circular motion. But as soon as their acidic particles become unsuitable for circular motion for some reason, then, exalting the ether of the first kind, they extinguish the red color and, leaving blue and yellow free, produce a green color, for example: when vitriolic, so-called oil (a substance surpassing all others in acidity), thickens in severe frost and its particles have very little circular motion, then a green color is born in it. Similarly, copper and iron, among other metals akin to acidic matters, which not only dissolve more readily in them themselves, but also break down in their vapors, showing mutual combination of particles of one kind, by joining together, losing convenience for circular motion, retain red ether with acidity; hence their solutions, crystals, and precipitates in pure acidic vitriolic oil[9] tend more towards a green color.

---

[8] Organic (vegetable or animal) alkali
[9] Strong sulfuric acid

I would like to present for the confirmation of this system all the examples from numerous experiments, especially those conducted by me in the study of multicolored glasses for mosaic art; I would like to explain everything that I have pondered over the colors for fifteen years, among my other works. But this requires, firstly, a very long time, longer than allowed for public discourse. Secondly, for clear interpretation of everything, it is necessary to present my entire system of physical chemistry, which my love for the Russian language[10], for the glorification of Russian heroes, and for the accurate investigation of the deeds of our homeland prevents me from accomplishing and communicating to the learned world.

Therefore, I now request that this expression of my thoughts on the origin of colors be accepted favorably and patiently awaited, if God deems it worthy, of my whole system. Especially do I present it to those who, while dealing with praise in one chemical practice, dare not raise their heads above coal and ash, so that they may not deem investigations into the causes and nature of primary particles, constituting bodies, from which colors and other properties of sensitive bodies arise, to be futile and sophistical. For knowledge of primary particles is necessary in physics only if the primary particles are necessary for the composition of sensitive bodies. For what purpose were many experiments conducted in physics and chemistry? For what purpose did great men toil and undergo dangerous trials in their lives? Was it only to gather a great multitude of different things and materials into a disorderly heap, to gaze at and marvel at their multitude, without reflecting on their arrangement and bringing them into order?

So when simple conjectures without any evidence and propositions subjected to inconvenient difficulties have served many for glory in the whole learned world, then I also expect from it that my system will be worthy of their attention. The importance of the topic will prompt it. A large part of the pleasures and delights in our lives depends on colors. The beauty of the human face, clothing, and other decorations and utensils, the pleasantness of various minerals and precious stones, the beauty of animals of various kinds, and finally all the radiance of the favorable and beautiful sun, all that it produces in its magnificence over blooming fields, in forests, and in seas – isn't all that worthy of our attention?

Having briefly presented my opinion on this difficult but joyful matter, appropriate to the present celebration, I turn to you, listeners, whose hearts are filled with joy which is not contained within them and overflows onto your faces and eyes. Your thoughts turn to past exclamations and celebrations during the days of Peter, now returned and multiplied by the divine blessing and happiness of the great Elizabeth, by the solemn commemoration of the most illustrious sovereigns and grand princes Peter and Paul. With your congratulations, listeners, and those of the general public, the Imperial Academy of Sciences offers to them, through an extraordinary public assembly, the humble expression of reverence and joy. Oh, how magnificent and splendid is the image we have of this reigning spring amidst our delights! An image of majesty, power, glory, and all virtues unmatched by our most noble Empress! An image of all benevolence, mutual love, and other great blessings of the blessed spouses, their imperial highnesses! An image of their beloved young heirs, the sweetest hope and expectation of our hearts! All your desires, listeners, and those of the whole homeland, echo ours. Oh, beautiful, dearest, and most beloved flower, grown from the noblest root in all of Europe, most illustrious sovereign, Grand Prince Pavel Petrovich, bloom amidst the abundance of the vast garden of the all-Russian state, renewed and fortified by the immortal labors of your great forefather, adorned with his praiseworthy virtues and divine blessings, by the zealously imitating legitimate heirs, worthy daughters of so great a father, our most gracious sovereign. Grow in the radiance of the timeless sun; delight us all with the fragrance of universal joy; gladden our eyes and hearts with the unfading beauty of your invaluable health; attain unimpeded full maturity; multiply the cherished fruits of the inheritance for the eternal pleasure of the homeland.

---

[10] Lomonosov references his works on Russian philology, poem "Peter the Great", and "Ancient Russian History".

## СЛОВО
## О ПРОИСХОЖДЕНИИ СВѢТА

### НОВУЮ ТЕОРІЮ
### О ЦВѢТАХЪ

ПРЕДСТАВЛЯЮЩЕЕ
ВЪ ПУБЛИЧНОМЪ СОБРАНІИ
ИМПЕРАТОРСКОЙ АКАДЕМІИ НАУКЪ
ІЮЛЯ 1 ДНЯ 1756 ГОДА
ГОВОРЕННОЕ
МИХАЙЛОМЪ ЛОМОНОСОВЫМЪ.

Печатано въ Санктпетербургѣ при Императорской Академіи Наукъ

1756

---

## ORATIO
## DE
## ORIGINE LVCIS
SISTENS
## NOVAM THEORIAM COLORVM,
IN PVBLICO CONVENTV
ACADEMIAE SCIENTIARVM IMPERIALIS
PETROPOLITANAE
PROPTER
NOMINIS FESTIVITATEM
SERENISSIMI PRINCIPIS
## MAGNI DVCIS
## PAVLI PETRIDAE
HABITA CALENDIS IVLIIS
ANNI cIɔ Iɔcc LVI.
A
*MICHAELE LOMONOSOW*
CONSILIARIO ACADEMICO.

EX ROSSICA IN LATINAM LINGVAM CONVERSA
a
GREGORIO KOSITZKI.

PETROPOLI,
Typis Academiae Scientiarum.

*Commentary* [by V.Shiltsev]:

This English translation of Mikhail Lomonosov's seminal work draws from its original Russian and Latin sources [1] (also Ref. [2], v.3, pp.315-344.). It is part of a series of English translations of Lomonosov's nine most significant scientific treatises, all of which Lomonosov himself compiled in the volume *titled Lomonosow Opera Academica*, intended for dissemination among European Academies. Seven other translations of these works can be found in Refs. [3-8]. Henry Leicester's book [9] includes translations of "Oration on the Origin of Light" together with "Oration on the Use of Chemistry" and "Meditations on the Solidity and Fluidity of Bodies". Unfortunately, these translations are found to be not fully satisfactory, as the meaning of the Russian original was sometimes lost or distorted due to the difficult 18th-century Russian of Lomonosov's original publication, which may not always be easily readable or understood even by contemporary Russian speakers. This rendition addresses the deficiencies of the H. Leicester's translation (which was used only for reference) and was initially composed with the aid of Google Translate and ChatGPT programs, providing a rough draft from both Russian and Latin versions of Lomonosov's "*Oration…*". However, machine translations often failed to capture the nuances and complexities of the original text, again, due to the challenging old Russian language of the original. Therefore, extensive reworking and improvements by the translator were indispensable to ensure the translation accurately conveys the intended meaning and is free from semantic deficiencies. For further exploration of the life and contributions of Mikhail Lomonosov (1711-1765) - the eminent Russian polymath and a towering figure of the European Enlightenment - readers are encouraged to consult various books [10, 11] and recent articles [12-18].

This "*Oration…*" was read by Lomonosov at the public meeting of the St.Petersburg Academy of Sciences on July 1, 1956, and published as a separate edition in 1757 in Russian, shortly followed by the 1758 Latin translation (with an equal print run of 400 copies each). The work garnered attention and commentary from scholars across Western Europe at the time, including Thomas Young, who referenced it in his own work on color theory – see Ref.[19], v.II, p.290.

Lomonosov's fascination with optics traces back to the early stages of his career, and his exploration of mosaic craftsmanship expanded this interest into the realm of color phenomena. Upon initiating experiments in colored glass preparation in the chemical laboratory (see, e.g., [18]) established for him in 1748, he delved deeply into this area of study. Through correspondence with Euler between 1750 and 1754, he not only undertook practical experiments but also developed a comprehensive theory of color. Expanding his corpuscular theory beyond the mechanics of material bodies, Lomonosov ventured into the kinetics of ether, intertwining his chemical and physical perspectives to formulate a rudimentary wave theory of light and color. This marked a significant departure from his earlier focus on theoretical physics and chemistry, which had waned around 1750. By 1756, he deemed his theories mature enough for presentation to the Academy.

Lomonosov opens his "*Oration…*" by positing that colors stem from light, prompting a rigorous examination of its origins and characteristics. Two prevailing schools of thought on the nature of light emerge: one championed by Huygens and Descartes, and the other tracing its lineage from Gassendi and gaining significance through the alignment and interpretation of Newton. Both perspectives posit ether as the medium of light transmission, diverging primarily in the manner by which its particles convey light through their motion. In a compelling argument, Lomonosov refutes the notion that light propagates via emanations of ether from luminous bodies, as posited by Gassendi and Newton, drawing upon his earlier work, "Reflections on the Causes of Heat and Cold," (Ref.[2], v.2, pp.57-61) wherein he theorized that heat is transmitted via the rotational motion of particles. He asserts that only the circular (rotational) motion of ether particles can be attributed as the true cause of light.

Regarding the genesis of colors, Lomonosov elucidates that the ether permeating space consists of spherical particles of varying diameters - large, medium, and small - each adorned with protrusions and indentations. According to Lomonosov's principle of "particle matching," only particles of identical diameter (termed "compatible") can interlock, while those of different diameters ("incompatible") cannot, thereby their motions do not affect each other. Lomonosov presents a nuanced framework wherein the arrangement of ether particles within the spaces between larger particles determines the emergence of distinct colors. The interplay of these particles with "primary matters" on the illuminated surface of objects orchestrates the formation of colors, with each type of ether particle corresponding to a specific color.

Lomonosov then postulates three types of ether particles: "ether of the first magnitude with salt, ether of the second magnitude with mercury, ether of the third magnitude with sulfur or combustible primary matter" and considers possibilities for the movement of two types of particles with a stationary third, or for the collective movement of all three types of particles. He concludes "…Finally, I find that from the first type of ether comes the color red, from the second - yellow, and from the third - blue. Other colors arise from the mixture of the first ones." In cases where all "primary matters" are present on the illuminated surface of the object, all three types of ether particles combine with them weakly, causing the rays to be reflected in the eye, and there all three colors mix into white.

In his discourse, Lomonosov aligns with E. Mariotte's assertion that Newton's decomposition of white light into seven colors should be confined to three. He reinforces his theory with illustrative examples and underscores the pivotal role of rotational motion in both heat generation and color manifestation. In conclusion, Lomonosov expresses a desire to expound upon his color theory further, contingent upon the development of his comprehensive system of physical chemistry. Despite his scholarly commitments to Russian history, Lomonosov remains hopeful that he will achieve this feat, concluding with customary praise for Empress Elizabeth and "the most illustrious sovereign Grand Duke Pavel Petrovich."

I would like to thank Prof. Robert Crease of SUNY, my long-term collaborator and co-author of several scholar papers on Mikhail Lomonosov, for the encouragement to translate Lomonosov's major works to English.